\documentclass[a4paper,12pt]{spieman}
\usepackage{tabularx}
\usepackage{setspace}
\usepackage{tocloft}
\usepackage{amsmath}
\usepackage{amssymb}

\usepackage{xspace}
\usepackage{graphicx}
\newcommand{\ignore}[1]{}
\providecommand{\ao}{}

\renewcommand{\ao}{adaptive optics (AO)\renewcommand{\ao}{AO\xspace}\renewcommand{\Ao}{AO\xspace}\xspace}
\newcommand{\Ao}{Adaptive optics (AO)\renewcommand{\ao}{AO\xspace}\renewcommand{\Ao}{AO\xspace}\xspace}
\newcommand{\wfs}{wavefront sensor (WFS)\renewcommand{\wfs}{WFS\xspace}\renewcommand{\wfss}{WFSs\xspace}\xspace}
\newcommand{\wfss}{wavefront sensors (WFSs)\renewcommand{\wfs}{WFS\xspace}\renewcommand{\wfss}{WFSs\xspace}\xspace}
\newcommand{\shwfs}{Shack-Hartmann \wfs (SHWFS)\renewcommand{\shwfs}{SHWFS\xspace}\xspace}
\newcommand{\dm}{deformable mirror (DM)\renewcommand{\dm}{DM\xspace}\renewcommand{\dms}{DMs\xspace}\renewcommand{\Dms}{DMs\xspace}\renewcommand{\Dm}{DM\xspace}\xspace}
\newcommand{\dms}{deformable mirrors (DMs)\renewcommand{\dm}{DM\xspace}\renewcommand{\dms}{DMs\xspace}\renewcommand{\Dms}{DMs\xspace}\renewcommand{\Dm}{DM\xspace}\xspace}
\newcommand{\Dms}{Deformable mirrors (DMs)\renewcommand{\dm}{DM\xspace}\renewcommand{\dms}{DMs\xspace}\renewcommand{\Dms}{DMs\xspace}\renewcommand{\Dm}{DM\xspace}\xspace}
\newcommand{\Dm}{Deformable mirror (DM)\renewcommand{\dm}{DM\xspace}\renewcommand{\dms}{DMs\xspace}\renewcommand{\Dms}{DMs\xspace}\renewcommand{\Dm}{DM\xspace}\xspace}

\newcommand{\lqg}{linear-quadratic-gaussian (LQG)\renewcommand{\lqg}{LQG\xspace}\xspace}
\newcommand{\shs}{Shack-Hartmann sensor (SHS)\renewcommand{\shs}{SHS\xspace}\renewcommand{\shss}{SHSs\xspace}\xspace}
\newcommand{\shss}{Shack-Hartmann sensors (SHSs)\renewcommand{\shs}{SHS\xspace}\renewcommand{\shss}{SHSs\xspace}\xspace}
\newcommand{\lgs}{laser guide star (LGS)\renewcommand{\lgs}{LGS\xspace}\renewcommand{\lgss}{LGSs\xspace}\xspace}
\newcommand{\lgss}{laser guide stars (LGSs)\renewcommand{\lgs}{LGS\xspace}\renewcommand{\lgss}{LGSs\xspace}\xspace}
\newcommand{\Ngs}{Natural guide star (NGS)\renewcommand{\ngs}{NGS\xspace}\renewcommand{\Ngs}{NGS\xspace}\renewcommand{\ngss}{NGSs\xspace}\xspace}
\newcommand{\ngs}{natural guide star (NGS)\renewcommand{\ngs}{NGS\xspace}\renewcommand{\Ngs}{NGS\xspace}\renewcommand{\ngss}{NGSs\xspace}\xspace}
\newcommand{\ngss}{natural guide stars (NGSs)\renewcommand{\ngs}{NGS\xspace}\renewcommand{\Ngs}{NGS\xspace}\renewcommand{\ngss}{NGSs\xspace}\xspace}
\newcommand{\mems}{Micro-Electro-Mechanical Systems (MEMS)\renewcommand{\mems}{MEMS\xspace}\xspace}
\newcommand{\snr}{signal to noise ratio (SNR)\renewcommand{\snr}{SNR\xspace}\xspace}
\newcommand{\Moao}{Multi-object \ao (MOAO)\renewcommand{\moao}{MOAO\xspace}\renewcommand{\Moao}{MOAO\xspace}\xspace}
\newcommand{\moao}{multi-object \ao (MOAO)\renewcommand{\moao}{MOAO\xspace}\renewcommand{\Moao}{MOAO\xspace}\xspace}
\newcommand{\mcao}{multi-conjugate adaptive optics (MCAO)\renewcommand{\mcao}{MCAO\xspace}\xspace}
\newcommand{\ltao}{laser tomographic \ao (LTAO)\renewcommand{\ltao}{LTAO\xspace}\xspace}
\newcommand{\cpu}{central processing unit (CPU)\renewcommand{\cpu}{CPU\xspace}\renewcommand{\cpus}{CPUs\xspace}\xspace}
\newcommand{\cpus}{central processing units (CPUs)\renewcommand{\cpu}{CPU\xspace}\renewcommand{\cpus}{CPUs\xspace}\xspace}
\newcommand{\psf}{point spread function (PSF)\renewcommand{\psf}{PSF\xspace}\renewcommand{\psfs}{PSFs\xspace}\renewcommand{\Psf}{PSF\xspace}\xspace}
\newcommand{\psfs}{point spread functions (PSFs)\renewcommand{\psf}{PSF\xspace}\renewcommand{\psfs}{PSFs\xspace}\renewcommand{\Psf}{PSF\xspace}\xspace}
\newcommand{\Psf}{Point spread function (PSF)\renewcommand{\psf}{PSF\xspace}\renewcommand{\psfs}{PSFs\xspace}\renewcommand{\Psf}{PSF\xspace}\xspace}
\newcommand{\fpga}{field programmable gate array (FPGA)\renewcommand{\fpga}{FPGA\xspace}\renewcommand{\fpgas}{FPGAs\xspace}\xspace}
\newcommand{\fpgas}{field programmable gate arrays (FPGAs)\renewcommand{\fpga}{FPGA\xspace}\renewcommand{\fpgas}{FPGAs\xspace}\xspace}
\newcommand{\sor}{successive over-relaxation (SOR)\renewcommand{\sor}{SOR\xspace}\xspace}
\newcommand{\fdpcg}{Fourier domain pre-conditioned gradient (FDPCG)\renewcommand{\fdpcg}{FDPCG\xspace}\xspace}
\newcommand{\map}{maximum a-posteriori (MAP)\renewcommand{\map}{MAP\xspace}\xspace}

\newcommand{\elt}{Extremely Large Telescope (ELT)\renewcommand{\elt}{ELT\xspace}\renewcommand{\elts}{ELTs\xspace}\renewcommand{\eelt}{European ELT (E-ELT)\renewcommand{\eelt}{E-ELT\xspace}\xspace}\xspace}

\newcommand{\elts}{Extremely Large Telescopes (ELTs)\renewcommand{\elt}{ELT\xspace}\renewcommand{\elts}{ELTs\xspace}\renewcommand{\eelt}{European ELT (E-ELT)\renewcommand{\eelt}{E-ELT\xspace}\xspace}\xspace}

\newcommand{\eelt}{European Extremely Large Telescope (E-ELT)\renewcommand{\eelt}{E-ELT\xspace}\renewcommand{\elt}{ELT\xspace}\renewcommand{\elts}{ELTs\xspace}\xspace}

\newcommand{\dugall}{Durham University generalised adaptive optics laser laboratory (DUGALL)\renewcommand{\dugall}{DUGALL\xspace}\xspace}
\newcommand{\fwhm}{full-width at half-maximum (FWHM)\renewcommand{\fwhm}{FWHM\xspace}\xspace}
\newcommand{\wht}{William Herschel Telescope (WHT)\renewcommand{\wht}{WHT\xspace}\xspace}
\newcommand{\emccd}{electron multiplying CCD (EMCCD)\renewcommand{\emccd}{EMCCD\xspace}\renewcommand{\emccds}{EMCCDs\xspace}\xspace}
\newcommand{\emccds}{electron multiplying CCDs (EMCCDs)\renewcommand{\emccd}{EMCCD\xspace}\renewcommand{\emccds}{EMCCDs\xspace}\xspace}
\newcommand{\dasp}{Durham \ao simulation platform (DASP)\renewcommand{\dasp}{DASP\xspace}\renewcommand{\thedasp}{DASP\xspace}\renewcommand{\Thedasp}{DASP\xspace}\xspace}
\newcommand{\thedasp}{the Durham \ao simulation platform (DASP)\renewcommand{\dasp}{DASP\xspace}\renewcommand{\thedasp}{DASP\xspace}\renewcommand{\Thedasp}{DASP\xspace}\xspace}
\newcommand{\Thedasp}{The Durham \ao simulation platform (DASP)\renewcommand{\dasp}{DASP\xspace}\renewcommand{\thedasp}{DASP\xspace}\renewcommand{\Thedasp}{DASP\xspace}\xspace}
\newcommand{\mpi}{Message Passing Interface (MPI)\renewcommand{\mpi}{MPI\xspace}\xspace}
\newcommand{\smp}{symmetric multi-processing (SMP)\renewcommand{\smp}{SMP\xspace}\xspace}
\newcommand{\svd}{singular value decomposition (SVD)\renewcommand{\svd}{SVD\xspace}\xspace}
\newcommand{\gpu}{graphics processing unit (GPU)\renewcommand{\gpu}{GPU\xspace}\renewcommand{\gpus}{GPUs\xspace}\xspace}
\newcommand{\gpus}{graphics processing units (GPUs)\renewcommand{\gpu}{GPU\xspace}\renewcommand{\gpus}{GPUs\xspace}\xspace}
\newcommand{\fft}{fast Fourier transform (FFT)\renewcommand{\fft}{FFT\xspace}\xspace}
\newcommand{\ifu}{integral field unit (IFU)\renewcommand{\ifu}{IFU\xspace}\xspace}
\newcommand{\darc}{the Durham \ao real-time controller (DARC)\renewcommand{\darc}{DARC\xspace}\renewcommand{\Darc}{DARC\xspace}\xspace}
\newcommand{\Darc}{The Durham \ao real-time controller (DARC)\renewcommand{\darc}{DARC\xspace}\renewcommand{\Darc}{DARC\xspace}\xspace}
\newcommand{\cots}{commercial off-the-shelf (COTS)\renewcommand{\cots}{COTS\xspace}\xspace}
\newcommand{\rtcp}{real-time control pipeline (RTCP)\renewcommand{\rtcp}{RTCP\xspace}\xspace}
\newcommand{\rms}{root-mean-square (RMS)\renewcommand{\rms}{RMS\xspace}\xspace}
\newcommand{\sFPDP}{serial Front Panel Data Port (sFPDP)\renewcommand{\sFPDP}{sFPDP\xspace}\xspace}
\newcommand{\wpu}{wavefront processing unit (WPU)\renewcommand{\wpu}{WPU\xspace}\xspace}

\newcommand{\rtcs}{real-time control system (RTCS)\renewcommand{\rtcs}{RTCS\xspace}\renewcommand{\rtcss}{RTCSs\xspace}\xspace}
\newcommand{\rtcss}{real-time control systems (RTCSs)\renewcommand{\rtcs}{RTCS\xspace}\renewcommand{\rtcss}{RTCSs\xspace}\xspace}
\newcommand{\eso}{European Southern Observatory (ESO)\renewcommand{\eso}{ESO\xspace}\renewcommand{\theeso}{ESO\xspace}\xspace}
\newcommand{\theeso}{\renewcommand{\theeso}{ESO\xspace}the \eso}
\newcommand{\scao}{single conjugate \ao (SCAO)\renewcommand{\scao}{SCAO\xspace}\renewcommand{\Scao}{SCAO\xspace}\xspace}
\newcommand{\Scao}{Single conjugate \ao (SCAO)\renewcommand{\scao}{SCAO\xspace}\renewcommand{\Scao}{SCAO\xspace}\xspace}
\newcommand{\glao}{ground layer \ao (GLAO)\renewcommand{\glao}{GLAO\xspace}\xspace}
\newcommand{\eagle}{ELT Adaptive optics for GaLaxy Evolution (EAGLE)\renewcommand{\eagle}{EAGLE\xspace}\xspace}
\newcommand{\maory}{multi-conjugate \ao relay for the \eelt (MAORY)\renewcommand{\maory}{MAORY\xspace}\xspace}
\newcommand{\muse}{Multi Unit Spectroscopic Explorer (MUSE)\renewcommand{\muse}{MUSE\xspace}\xspace}
\newcommand{\vlt}{Very Large Telescope (VLT)\renewcommand{\vlt}{VLT\xspace}\xspace}
\newcommand{\ccd}{CCD\xspace}
\newcommand{\tmt}{Thirty Metre Telescope (TMT)\renewcommand{\tmt}{TMT\xspace}\xspace}
\newcommand{\xao}{eXtreme \ao (XAO)\renewcommand{\xao}{XAO\xspace}\xspace}

\newcommand{\vla}{Very Large Array (VLA)\renewcommand{\vla}{VLA\xspace}\xspace}
\newcommand{\jwst}{{\em James Webb Space Telescope} \citep[JWST,][]{jwst}\renewcommand{\jwst}{{\em JWST}\xspace}\xspace}
\newcommand{\hst}{{\em Hubble Space Telescope (HST)}\renewcommand{\hst}{{\em HST}\xspace}\xspace}
\newcommand{\ifss}{integral-field spectrographs (IFSs)\renewcommand{\ifss}{IFSs\xspace}\renewcommand{\ifs}{IFS\xspace}\xspace}
\newcommand{\ifs}{integral-field spectrograph (IFS)\renewcommand{\ifss}{IFSs\xspace}\renewcommand{\ifs}{IFS\xspace}\xspace}
\newcommand{\ifus}{integral field units (IFUs)\renewcommand{\ifus}{IFUs\xspace}\xspace}
\newcommand{\mos}{multi-object spectrograph (MOS)\renewcommand{\mos}{MOS\xspace}\xspace}
\newcommand{\goodss}{Great Observatories Origins Deep Survey (GOODS)-S\renewcommand{\goodss}{GOODS-S\xspace}\xspace}
\newcommand{\goods}{Great Observatories Origins Deep Survey (GOODS)\renewcommand{\goods}{GOODS\xspace}\xspace}
\newcommand{\scmos}{scientific CMOS (sCMOS)\renewcommand{\scmos}{sCMOS\xspace}\xspace}
\newcommand{\aof}{Adaptive Optics Facility (AOF)\renewcommand{\aof}{AOF\xspace}\xspace}
\newcommand{\dsp}{digital signal processor (DSP)\renewcommand{\dsp}{DSP\xspace}\renewcommand{\dsps}{DSPs\xspace}\xspace}
\newcommand{\dsps}{digital signal processors (DSPs)\renewcommand{\dsp}{DSP\xspace}\renewcommand{\dsps}{DSPs\xspace}\xspace}
\newcommand{\capi}{Coherent Accelerator Processor Interface (CAPI)\renewcommand{\capi}{CAPI\xspace}\xspace}
\newcommand{\qe}{quantum efficiency (QE)\renewcommand{\qe}{QE\xspace}\xspace}

\newcommand{\mnras}{MNRAS}

\title{Analysis of EMCCD and sCMOS readout noise models for Shack-Hartmann
wavefront sensor accuracy}

\author{A.\ G.\ Basden,\supscr{a}}
\affiliation{\supscrsm{a}Department of Physics, South Road, Durham, DH1 3LE, UK}

\cftpagenumbersoff{figure}
\cftpagenumbersoff{table}

\begin{document}
\maketitle
\begin{abstract}
In recent years, detectors with sub-electron readout noise have been
used very effectively in astronomical adaptive optics systems.  Here,
we compare readout noise models for the two key faint flux level
detector technologies that are commonly used: EMCCD and scientific
CMOS (sCMOS) detectors.  We find that in almost all situations, EMCCD
technology is advantageous, and that the commonly used simplified
model for EMCCD readout is appropriate.  We also find that the commonly used
simple models for sCMOS readout noise are optimistic, and recommend
that a proper treatment of the sCMOS rms readout noise probability
distribution should be considered during instrument performance
modelling and development.  
\end{abstract}

{\noindent \footnotesize{\bf Address all correspondence to}: Alastair
  Basden, Durham University, Department of Physics, South Road,
  Durham, DH1 3LE, UK; Tel: +44 191 3342229; E-mail:
  \linkable{a.g.basden@durham.ac.uk} }

%\begin{spacing}{2}   % use double spacing for rest of manuscript

\section{Introduction}
Within the last decade, the use of optical detector arrays with
sub-electron readout noise has become common for wavefront sensors on
astronomical \ao systems.  The majority of these detectors have used
\emccd technology \cite{emccd}, for example as used by the CANARY
wide-field \ao demonstrator \cite{canaryshort} on the \wht and the
SPHERE \xao system \cite{sphere} on the \vlt.  However, \scmos
technology \cite{scmos} is now also offering sub-electron readout
noise, and is a potential alternative to \emccds, particularly
when larger detector arrays are required, for example for \lgs \wfss
for \elt-scale instruments.  An \scmos camera has been
used on-sky by CANARY during \lgs commissioning.

\emccd and \scmos detectors have different readout noise
characteristics.  The relative effect of different readout noise
models on \shs \wfs images, and the corresponding wavefront slope
estimation accuracy, has not previously been studied in depth.

\subsection{EMCCD readout noise}
\emccds work on the principal of impact ionisation, where as the
signal in a given pixel (electrons) are transferred along a many-stage
multiplication register, there is a small probability (typically of
order 1\%, $p=0.01$) that each photo-electron will generate an
additional electron.  These registers are many hundreds of elements
long, and so a large mean multiplication (or gain) can be achieved,
equal to $(1+p)^n$ where $n$ is the number of stages.  Unfortunately,
this multiplication process is stochastic, and for a given number of
input photons in a given pixel, there is a wide range in the possible
measured \emccd output value \cite{basden1}, in addition to photon
shot noise which is always present.  Typically, a gain of order
500-1000 is used for astronomical \ao systems.  After the signal has
been multiplied in this way, it is then read out of the detector and
digitised, introducing readout noise to the signal.  This readout
noise is dependent on readout speed, and typically has a rms of about
50 electrons for an \emccd operated at high frame rates.  We ignore
thermal noise, since \emccds typically operate at high frame rates,
and are usually cooled to temperatures of around 220~K in commercial
camera models.

When modelling the impact of detector performance on instrument
designs, the combination of these sources of uncertainty leads to
increased complexity.  Therefore, simplified models are often used
(for example in Ref.~\citenum{basden18}): typically, when modelling an \emccd,
the detector \qe is halved (i.e.\ the input flux is halved) as an
approximation of the effect of the stochastic gain mechanism, and a
readout of around 0.1~electrons is assumed (the 
true readout noise divided by the gain).  Here, we investigate the
effect of these assumptions.

\subsection{sCMOS readout noise}
\begin{figure}\begin{center}
\includegraphics[width=0.6\linewidth]{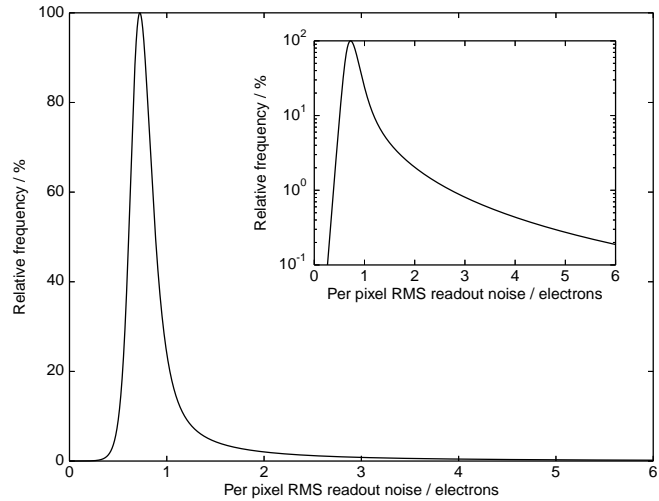}
\end{center}
\caption{The probability distribution for rms readout noise of
  individual sCMOS detector pixels, scaled to the frequency of the
  modal average.  Inset is shown the distribution
  on a logarithmic scale.  }
\label{fig:scmosProbDist}
\end{figure}

An \scmos detector is an active pixel sensor, with each pixel having
its own individual readout, rather than a single, or small number of,
readout ports in the case of a \ccd.  Each \scmos pixel will therefore
have an associated readout noise level, which will differ from other
readouts due to manufacturing imperfections, etc.  Additionally, the
readout noise introduced at each pixel will also vary with each frame
readout, i.e.\ the readout noise of a given pixel has some rms value,
with all pixels forming a rms readout noise probability distribution.
Therefore, manufacturers of \scmos cameras usually quote the median
rms readout of the device, which is at the level of 0.8 electrons for
the best current cameras.  As with an \emccd, this level will be
dependent on readout speed, which is generally not user-selectable for
current commercial \scmos cameras.  Fig.~\ref{fig:scmosProbDist} shows
the a histogram of variation of individual pixel rms readout for a
typical \scmos device \cite{pcoedge42}.

Instrument modelling of \scmos detectors has to date typically used a
single rms readout value for all pixels (see for example
Ref.~\citenum{basden16}), and readout noise is often
described using a single (unspecified) parameter, for example
Ref.~\citenum{scmosEval}.  However, this can lead to an overestimation of
instrument performance, since the occasional pixels with far greater
readout noise are not modelled.

\subsection{Accurate readout noise modelling for Shack Hartmann
  wavefront sensors}
This paper seeks to investigate the effect of accurate readout noise
models on the performance of Shack Hartmann wavefront sensors commonly
used for astronomical \ao systems.  In \S2, we describe the models
used, our performance verification and the implemented tests.  In \S3,
we discuss our findings and summarise the results.  We conclude in
\S4.

\section{Modelling readout noise in Shack Hartmann wavefront sensors}
To investigate the effect of sensor readout noise characteristics on
Shack-Hartmann wavefront sensor performance, we perform Monte-Carlo
simulations of a single Shack-Hartmann sub-aperture, investigating
different spot sizes and different sub-aperture sizes (i.e.\ number
of pixels) for a range of input flux signal levels.  Our procedure,
following Ref.~\citenum{basden18}, is as follows:
\begin{enumerate}
\item A  noiseless sub-aperture spot is generated at a random
  position, and the centre of gravity calculated ($S_\textrm{x:true}$,
  $S_\textrm{y:true}$ for the x and y position respectively).
\item Random photon shot noise is introduced across the sub-aperture.
\item Detector readout is modelled (see \S\ref{sect:emccdModel} and
  \S\ref{sect:scmosModel}).
\item The spot position is estimated using a centre of gravity
  algorithm ($S_\textrm{x:estimated}$, $S_\textrm{y:estimated}$ for
  the x and y position respectively).
\item Steps 1--4 are repeated many ($N$) times.
\item The performance metric, $R$, is calculated.
\end{enumerate}
The performance metric is given by
\begin{equation}
R = \frac{\sum_{m=1}^N
    \sqrt{\left(S_\textrm{x:true}(m)-S_\textrm{x:estimated}(m)
      \right)^2 +
	\left(S_\textrm{y:true}(m)-S_\textrm{y:estimated}(m)
      	\right)^2 
 }}
{N}
\label{eq:metric}
\end{equation}
where $S(m)$ is the $m^{th}$ individual slope measurement (x or y,
true or estimated) of $N$ Monte-Carlo measurements (typically ten
thousand).  Essentially, this is the mean distance of the estimated
position from the true position.  We refer to this interchangeably as
the slope error (on figure axes), and as the slope estimation
accuracy.

We use an Airy disk for the noiseless sub-aperture spot, the width of which is
a parameter we investigate (to allow for performance estimates with
different pixel scales and seeing conditions), which we define here as
the diameter of the first Airy minimum in pixels.  When processing the
noisy images to compute the spot position, different background levels
are subtracted to enable investigation of optimum background
subtraction.  The background level resulting in lowest slope error is
then used.

Signal levels from 20 photons per sub-aperture (below what would be
used effectively on-sky) to 1000 photons per sub-aperture (approaching
a high light level condition) are used.  We assume 100\% \qe for the
detectors to simplify data analysis, except for the simplified
  \emccd model where the excess noise factor means that effective \qe
  is 50\% \cite{excessNoiseFactor,2006SPIE.6276E..1FD}.  In practise, the \qe of a
back-illuminated \emccd can approach 95\%, while second generation
\scmos detectors have a \qe greater than 70\%.  By scaling flux levels
by the relevant \qe (as we do in Fig.~\ref{fig:scaledForQe} to provide
an example), a reader can evaluate detector performance for their
particular image sensor.

Unless stated otherwise, we assume here a spot size of diameter 2
pixels (Airy ring minima), and a signal level of 50 photons per
sub-aperture.  We investigate these parameters, and the number of
pixels within a sub-aperture.

\subsection{EMCCD models}
\label{sect:emccdModel}
We introduce three models for \emccd technology readout noise:
\begin{enumerate}
\item EMCCD Simple:  The simple model, involving halving the effective
  detector quantum efficiency, and using a readout noise of 0.1
  electrons rms, normally distributed.
\item EMCCD Stochastic: A full stochastic Monte-Carlo electron multiplication
  process is modelled, with the photo-electrons from each pixel being
  propagated through the multiplication register, with a small, random
  probability of being multiplied at each stage.  A readout noise of
  50 electrons rms is then applied.
\item EMCCD Distribution: The \emccd output is obtained from the
  probability distribution given by Eq.~\ref{eq:emprob}, and a readout
  noise of 50 electrons rms is then applied.
\end{enumerate}
For the stochastic and probability distribution models, we use a mean
gain of 500 (unless otherwise stated), with 520 multiplication stages,
and therefore a probability of about 1.2\% of a new electron being
generated at each stage, for each input electron.  We also do not
investigate other readout noises which could be introduced at
different detector readout speeds.  To achieve the same performance at
other readout noise levels, the \emccd gain could be altered.

The probability distribution for \emccd output is given by
Eq.~\ref{eq:emprob}, taken from Ref.~\citenum{basden1}.  Additionally,
here we also introduce an approximation for this distribution at
higher light levels (e.g.\ for greater than 50 input photo-electrons):
\begin{equation}
p(x) = \frac{x^{n-1}\exp\left(-x/g\right)}{g^n(n-1)!}
\label{eq:emprob}
\end{equation}

And our high light level approximation
\begin{equation*}
p(x) =
\frac{\exp\left(-\left(x-g\left(n-1\right)\right)^2/\left(2g\sqrt{n}\right)\right)}{\sqrt{2\pi g^{2}n}}
\end{equation*}

where $n$ is the number of input photo-electrons, $g$ is the mean gain
and $x$ is the output of the probability distribution.  We use the
high light level approximation for input signal levels greater than 50
photo-electrons.

\subsubsection{Thresholding schemes}
We also investigate a thresholding scheme for \emccd output data, as
introduced by Ref.~\citenum{basden1}.  In particular, we use the Poisson
Probability (PP) scheme.  This concept involves taking the \emccd
output, dividing by the mean gain, and placing it into non-uniformly
spaced bins, with the $n^\mathrm{th}$ bin being interpreted as $n$ detected
photo-electrons.  The positions (thresholds) of bin boundaries that we use are given
by Ref.~\citenum{basden1}, placed where the probability of obtaining a
given output signal for a light level of $n$ photons and $n+1$ photons
is equal.

This thresholding scheme is non-linear, and as a result does not
provide a calibrated flux measurement.  Application of a photometric
correction is therefore also investigated here, as given in
Ref.~\citenum{basden1}.  We note that this scheme is far from perfect,
as identified in Ref.~\citenum{2006SPIE.6276E..1FD}, and so seek only
to investigate whether performance improvements are possible when
using it.

\subsection{sCMOS models}
\label{sect:scmosModel}
We also introduce a number of models for \scmos readout noise:
\begin{enumerate}
\item \scmos Median: All pixels have the same rms readout noise, normally distributed, equal to the
  manufacturer quoted median readout noise.
\item \scmos Mean: All pixels have the same rms readout noise, normally
  distributed, equal to the manufacturer quoted rms readout noise.
\item A different rms readout noise for each pixel following the
  probability distribution in Fig.~\ref{fig:scmosProbDist} (Eq.~\ref{eq:scmosprob}):
\begin{enumerate}
\item \scmos Fixed: We investigate 10 different sub-apertures, each
  following this probability distribution for readout noise, with an
  individual pixel's rms readout noise held constant over the entire
  Monte-Carlo simulation.
\item \scmos Random: We also investigate performance when the readout
  noise of pixels within a sub-aperture are changed each iteration
  (obeying the probability distribution), to get a feel for what
  ``overall'' performance would be like (i.e.\ many sub-apertures on
  the detector).  Effectively, we are sampling many different
  sub-apertures and obtaining a mean expected performance metric.  
\label{item:random}
\end{enumerate}
\end{enumerate}
It should be noted that in the cases where rms readout noise follows
the probability distribution Eq.~\ref{eq:scmosprob}, some pixels will
have a much larger rms readout noise than others, and thus will have a
negative effect on centroid estimates. We use a probability
distribution that closely matches manufacturer data\cite{pcoedge42},
given by:
\begin{equation}
f=\frac{1}{N}\left(\tanh\left(10x-6.5\right)+1\right)\times\left(\frac{1}{x^{10}+0.1}+\frac{1}{2\left(x^4+0.1\right)}+\frac{1}{2\left(x^2+0.1\right)}\right)
\label{eq:scmosprob}
\end{equation}
where $f$ is the probability of a given pixel having readout noise
$x$, and $N$ is the normalisation factor.
Fig.~\ref{fig:scmosProbDist} shows this distribution.  We assume a slow-scan
readout scheme to generate this probability distribution: a fast-scan
readout would introduce more noise, shifting the distribution.

\begin{figure}\begin{center}
\includegraphics[width=0.6\linewidth]{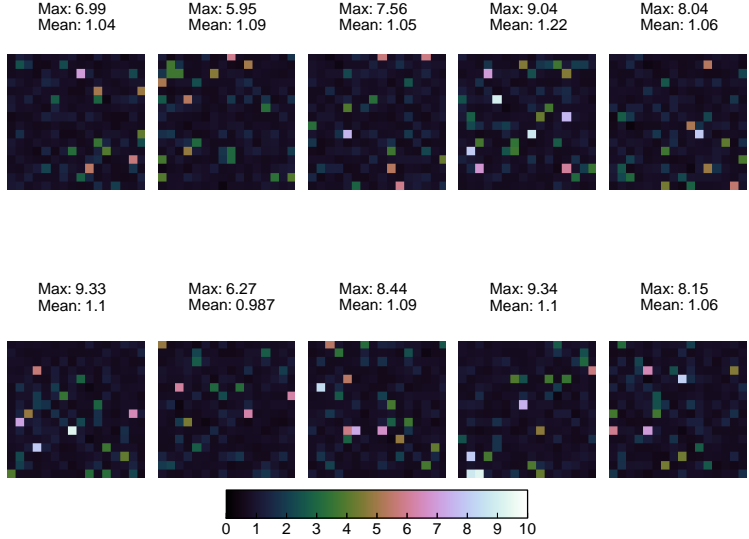}
\end{center}
\caption{The per-pixel rms readout noise for the 10 sample noise
  patterns used here for $16\times16$ pixel sub-apertures.  Each value
  represents the rms readout noise of that particular pixel, which is
  then randomly sampled from a Gaussian distribution every Monte-Carlo
  iteration.}
\label{fig:noisepattern}
\end{figure}

The random nature of pixel rms readout noise obtained from this
distribution means that some sub-apertures will behave very well (with
low noise throughout), while others will contain one or more noisy
pixels, particularly for larger sub-apertures.  To get some feel for
this effect, we randomly generate noise patterns for 10 different
sub-apertures, which are then used throughout the simulations (and for
interest are shown in Fig.~\ref{fig:noisepattern} for the $16\times16$
pixel case).  Additionally, to get a better estimate for the {\em
  mean} performance of the \scmos detector, we also include results
where a new rms readout noise pattern was obtained every Monte-Carlo
iteration (using the probability distribution,
Fig.~\ref{fig:scmosProbDist}), the ``\scmos Random'' model.  It is
important to note that these readout noise patterns are {\em not} a
static offset added to the image.  Rather, they represent the rms
readout noise of the individual pixels; for each Monte-Carlo
iteration, this rms value is used to generate the particular number of
noise electrons introduced, randomly distributed in a Gaussian
distribution with a standard deviation equal to the rms.

\section{Implications for instrumental modelling of low-noise detectors}
\begin{figure}\begin{center}
\includegraphics[width=0.6\linewidth]{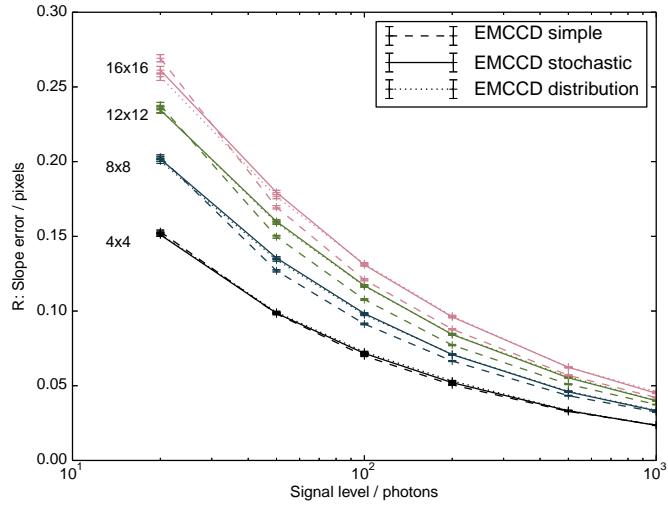}
\end{center}
\caption{A figure showing slope error as a function of signal level
  for the different EMCCD readout models, as given in the legend.  The
  groups of lines (differentiated by colour or shade) represent different
  sub-aperture sizes, as given by the annotations on the figure.  This
  figure is for a narrow spot diameter (Airy minimum separated by 2
  pixels).}
\label{fig:emccdres1}
\end{figure}

\begin{figure}\begin{center}
\includegraphics[width=0.6\linewidth]{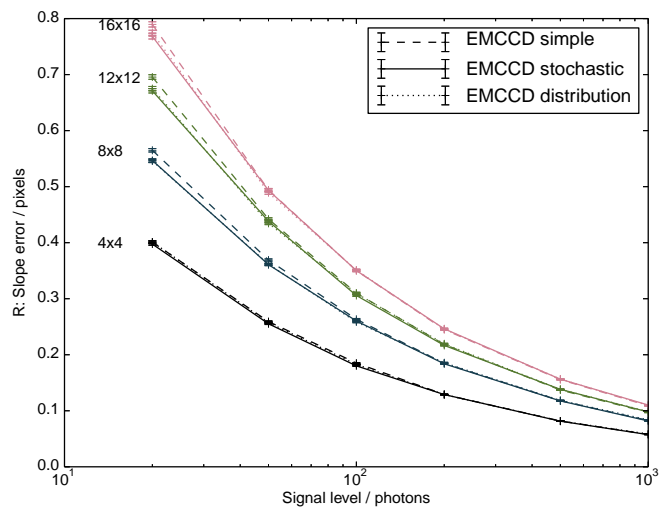}
\end{center}
\caption{A figure showing slope error as a function of signal level
  for the different EMCCD readout models, as given in the legend.  The
  groups of lines (differentiated by colour or shade) represent different
  sub-aperture sizes, as given by the annotations on the figure.  This
  figure is for a wide spot diameter (Airy minimum separated by 10
  pixels).}
\label{fig:emccdres5}
\end{figure}

The first question that we seek to answer is the appropriateness of
using the simplified \emccd model for instrument design decisions.  
Fig.~\ref{fig:emccdres1} and Fig.~\ref{fig:emccdres5} show slope error
as a function of signal level for different sub-aperture sizes, when
the different \emccd readout models are used.  It can be seen that the
probability distribution model agrees very closely with full
stochastic model.

At intermediate flux levels, the simple model underestimates slope
error slightly for small spot sizes (Fig.~\ref{fig:emccdres1}), while there is a
slight overestimation of slope error for larger spot sizes
(Fig.~\ref{fig:emccdres5}).  However, the difference between the simple
model and stochastic model are small, and unlikely to be a dominant
source of error for \ao instrument models.  We therefore recommend
that it is appropriate to use the simple \emccd model during \ao
system analysis and design.    

For astrometry, the case is not so simple.  Here, the difference in
spot position determination accuracy between the models may be more
significant.  Therefore we recommend that design studies for
astrometric instruments should at least investigate a full \emccd
stochastic model (or probability distribution model), rather than
assuming that the simple model is accurate enough.  We discuss this
further in \S\ref{sect:astrometry}.

\subsection{EMCCD gain}
\begin{figure}\begin{center}
\includegraphics[width=0.6\linewidth]{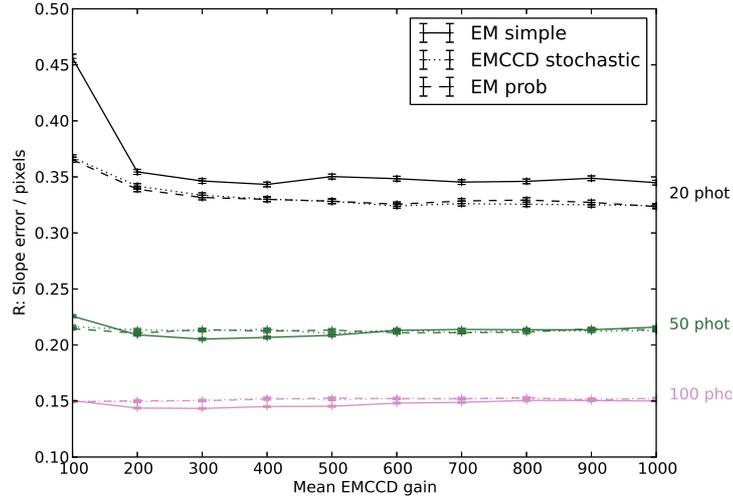}
\end{center}
\caption{A figure showing slope estimation error as a function of
  EMCCD mean gain, for different EMCCD readout models (given in the
  legend) at different light levels (given on the graph, in photons
  per sub-aperture), for a $8\times8$ pixel sub-aperture, with a spot
  of diameter 4 pixels (first Airy minimum).}
\label{fig:plotgain}
\end{figure}

Throughout our modelling, we have used a mean \emccd gain of 500,
which is close to the value that we frequently use on-sky with CANARY.
However, Fig.~\ref{fig:plotgain} also shows slope estimation error at
different levels of mean gain, for different light levels.  It is
clear here, that at the lowest light levels
performance predicted by the ``\emccd Simple'' model is worse than that of
other models.  We note that at 100 photons
per sub-aperture (and at higher light levels), the ``\emccd simple'' model
is optimistic.  We also note that the ``\emccd Stochastic'' and
``\emccd Distribution'' models give almost identical performance.

\subsection{Impact of thresholding schemes}
\begin{figure}\begin{center}
\includegraphics[width=0.6\linewidth]{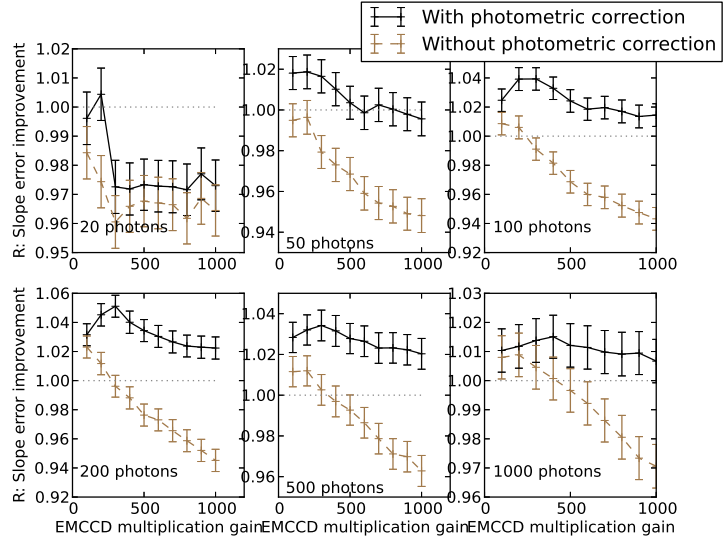}
\end{center}
\caption{The improvement in slope estimation accuracy resulting from
  application of a thresholding scheme as a function of
  \emccd gain at different signal levels (as given within the plots),
  compared to a stochastic multiplication gain model.  An improvement
  is signified by a value greater than unity (the dotted line at unity
  refers to performance without thresholding).  Results with, and
  without, photometric correction are given, and the sub-aperture size
  is $8\times8$ pixels, for a spot of diameter 4 pixels (first Airy minimum).}
\label{fig:emccdpp}
\end{figure}

Fig.~\ref{fig:emccdpp} shows the improvement in slope error brought
about by thresholding of the \emccd output for different signal
levels, as a function of \emccd gain, when compared with an
unprocessed stochastic multiplication model.  It can be seen that by
using the thresholding scheme and applying the photometric correction, a
reduction in slope estimation error is achievable, reducing the error
by up to 5\% under certain signal level conditions.  We note that the
photometric correction is necessary: applying thresholding without
this correction results in poorer performance.  The reduction in slope
error is at best about 5\%, and at the lowest light levels performance
is worse, and therefore we recommend that further investigation is
required for a given situation (sub-aperture size, spot size, etc)
before this strategy should be considered.

\subsection{sCMOS model implications}
The parameter most commonly given for \scmos readout noise by camera
manufacturers is the median value, which is as low as
0.8~photoelectrons for second generation devices.  The \rms readout
noise is also sometimes given, with typical values around
1.1~photoelectrons.  For instrument design studies, it can be tempting
to use either of these values, or something in between, when modelling
\scmos detectors, for example Ref.~\citenum{basden16} use a value of
1~photoelectron as representative of \scmos readout noise. 

Here, we compare slope estimation accuracy using both the typical median and
mean values, and also using models with inter-pixel variation in
readout noise, following the distribution given in
Fig.~\ref{fig:scmosProbDist}.  This probability distribution gives a
median readout noise of 0.8~photoelectrons, and a mean of 1.08.  

\begin{figure}\begin{center}
\includegraphics[width=0.6\linewidth]{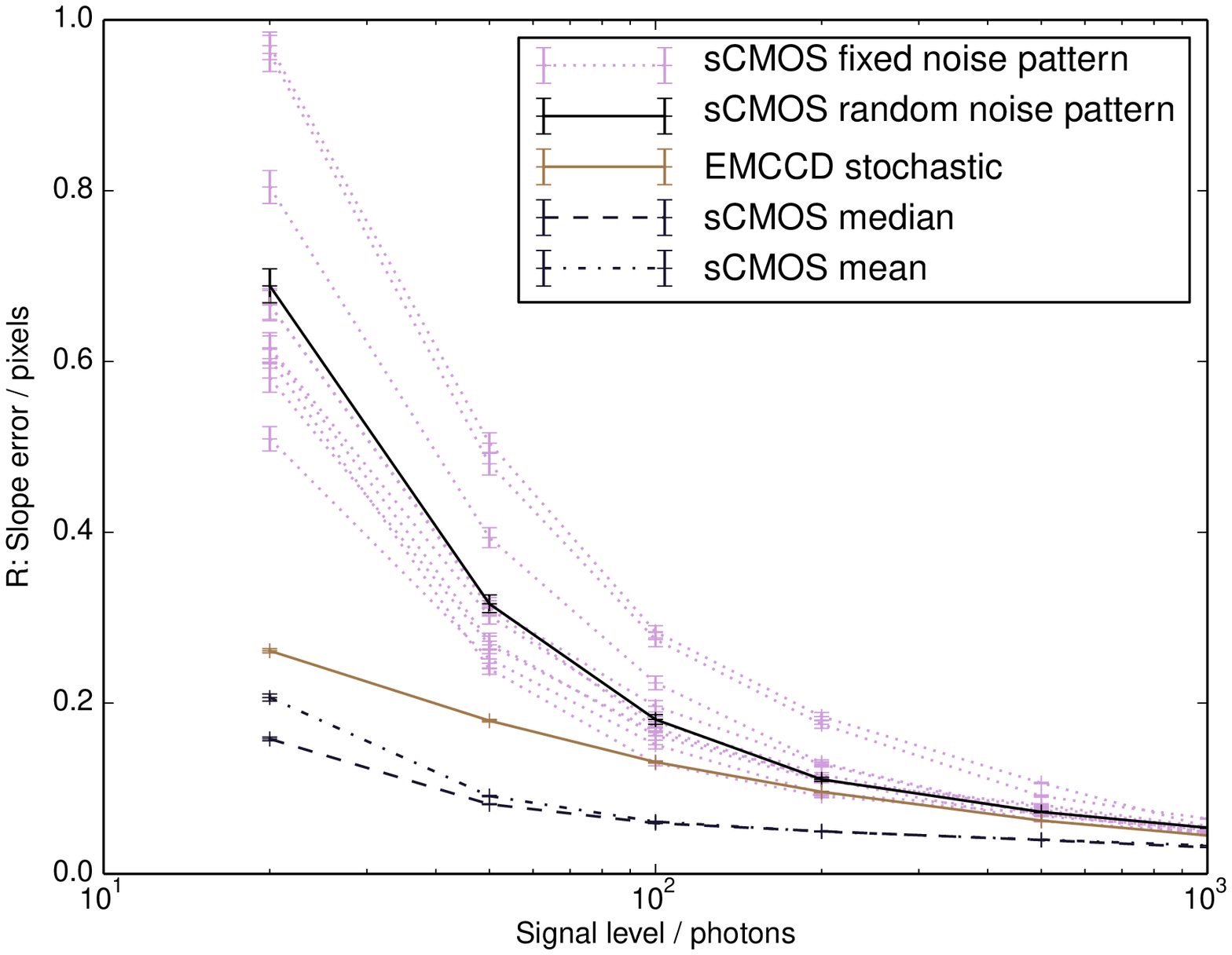}
\end{center}
\caption{A figure showing slope estimation error as a function of
  signal level for different detector readout models, for a
  sub-aperture with $16\times16$ pixels.}
\label{fig:scmosres16}
\end{figure}

Fig.~\ref{fig:scmosres16} shows slope estimation accuracy comparing
these different models when large sub-apertures ($16\times16$ pixels)
are used.  The \emccd stochastic model performance is also shown for
comparison.  It is interesting to note that using the median and mean
\scmos models provides significantly better performance than the
\emccd model.  At first sight, if one of these simple \scmos models is
used during instrument development it will appear that \scmos
technology is more appropriate for Shack-Hartmann wavefront sensing
than \emccd technology.  However, once the probability distribution
for readout noise is taken into account, this is clearly no longer the
case.  As Fig.~\ref{fig:scmosres16} shows, true \scmos performance is
significantly worse than that predicted using the simple models.  

\subsubsection{The spread of sub-aperture performance}
The ten curves for sub-apertures with different fixed readout noise patterns in
Fig.~\ref{fig:scmosres16} show a significant spread in slope error.
This is because some of these sub-apertures are ``unlucky'' (Fig.~\ref{fig:noisepattern}), in that
they contain one or more pixels with readout noise in the tail of the
probability distribution (Fig.~\ref{fig:scmosProbDist}).  Even the
``lucky'' sub-apertures, which yield lowest slope error, still have
performance significantly worse than simple readout models predict,
and still significantly worse than \emccd performance.  This is
because the pixels within these sub-apertures still have a range of
readout noise levels (the highest noise pixel in the best sub-aperture
having a readout noise of 5.95~electrons, and the highest noise pixel
in the worst sub-aperture having a readout noise of 9.34~electrons).  

To get an idea of ``average'' expected performance using a \scmos
detector, the ``\scmos Random'' model was used: every
Monte-Carlo iteration, each pixel is assigned a new rms readout noise
from the probability distribution.  This rms readout noise is then
used to obtain the number of readout electrons introduced that
iteration, using a Gaussian distribution with standard deviation equal
to the rms.  In effect, this allows us to sample average performance
over a large number of sub-apertures, and results are given by the
``\scmos random noise pattern'' curve in Fig.~\ref{fig:scmosres16}.  It can
be seen here that this offers significantly worse slope estimation
accuracy than either the \emccd or simple \scmos models.  

Currently available \scmos detectors all have large pixel counts.
Therefore, for applications requiring low order wavefront sensing,
where fewer pixels are required, it may be possible to select an area
of the \scmos detector where rms readout noise is generally low.
However, this will be device dependent, and we do not consider it
further here.

\subsubsection{Performance dependence on sub-aperture size}
\begin{figure}\begin{center}
\includegraphics[width=0.6\linewidth]{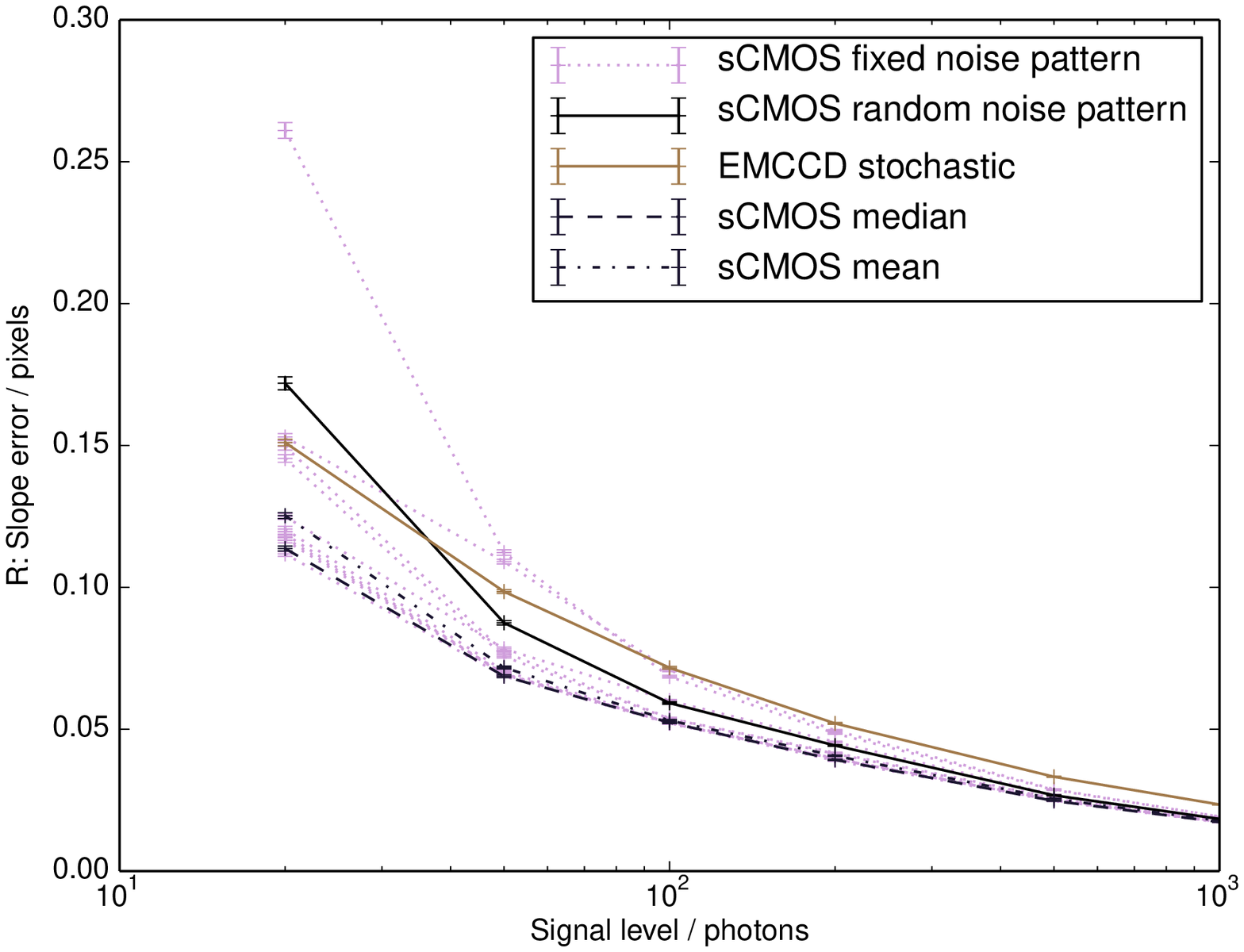}
\end{center}
\caption{A figure showing slope estimation error as a function of
  signal level for different detector readout models, for a
  sub-aperture with $4\times4$ pixels.}
\label{fig:scmosres4}
\end{figure}

Fig.~\ref{fig:scmosres4} shows slope estimation accuracy for different
detector readout models on a $4\times4$ pixel sub-aperture.  For all
but the lowest light levels, the ``average'' expected performance
using the \scmos detector (the ``\scmos Random'' model) is better than
that of the \emccd.  It is interesting to note that some sub-apertures
are ``lucky'', with performance at the level of that predicted by
simple \scmos models (i.e.\ constant readout noise equal to median or
mean).  This is because, with far fewer pixels, there is a higher
probability that all pixels within a sub-aperture can avoid the tail
of the probability distribution.  Of the 10 sub-aperture readout noise
patterns used, the maximum rms readout noise varied between 0.94 (for
the best sub-aperture) and 7 electrons (for the worst).  The mean rms
values ranged from 0.72 to 1.3 electrons.

\begin{figure}\begin{center}
\includegraphics[width=0.6\linewidth]{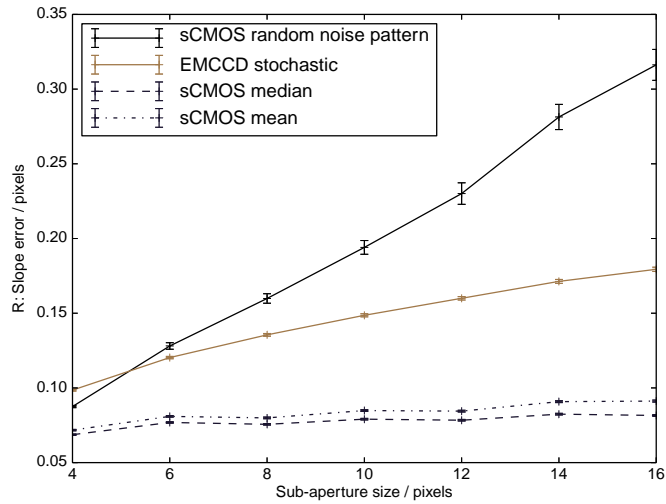}
\end{center}
\caption{A figure showing slope estimation error as a function of
  sub-aperture size for different detector readout models as given in
  the legend.  The sub-aperture size refers to the linear dimension,
  i.e.\ the square root of the total number of pixels within a
  sub-aperture.}
\label{fig:scmosnsubx}
\end{figure}

Fig.~\ref{fig:scmosnsubx} shows slope estimation accuracy as a
function of sub-aperture size.  It can be seen that \emccd performance
is better than \scmos performance for sub-aperture sizes equal to and
greater than $6\times6$ pixels.  For comparison, the simple \scmos
model results are also provided, and show that performance will be
greatly overestimated if these models are used.

Therefore, we recommend that proper models of \scmos readout noise
should always be used when modelling instrument performance.

\subsubsection{Performance dependence on spot size}
\begin{figure}\begin{center}
\includegraphics[width=0.6\linewidth]{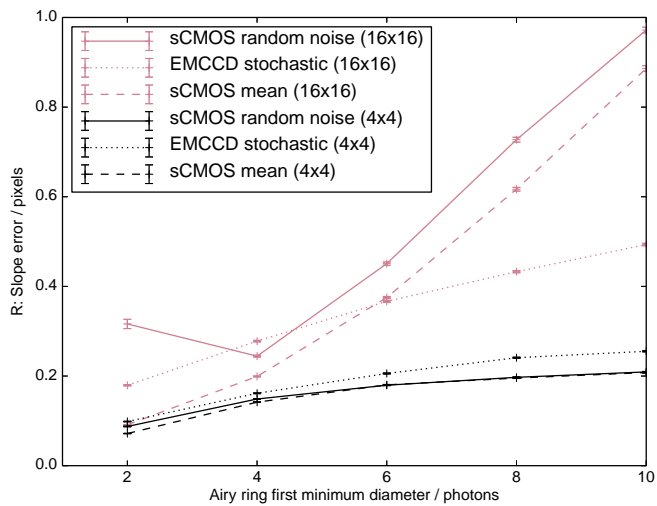}
\end{center}
\caption{A figure showing slope estimation error as a function of
  spot diameter (Airy ring first minimum diameter) for different
  detector readout models.}
\label{fig:plotdiam}
\end{figure}

Fig.~\ref{fig:plotdiam} shows slope estimation accuracy as a function of
Shack-Hartmann spot size.  For smaller sub-apertures, \scmos
performance is better than \emccd performance.  However, for larger
sub-apertures, \emccd performance is generally better, particularly as
spot size increases (with the available flux being spread over more
pixels).  We note that performance of \scmos technology predicted
using the full noise distribution model is always significantly worse
than performance predicted using a simple (constant rms readout noise)
model for larger sub-apertures.

\subsubsection{A simple model for sCMOS readout noise}
We have established that using the mean or median \scmos rms readout
noise when estimating instrumental performance is optimistic.
Unfortunately, using a full probability distribution will introduce
additional complexity to instrumental modelling, and increase the
parameter space that requires exploration, in part due to the need to
randomly sample different parts of the probability distribution (to
sample different areas of a detector) to obtain an average expected
performance.  Therefore, if a single-parameter model \scmos readout
noise can be obtained, this will greatly simplify instrumental
modelling.

\begin{figure}\begin{center}
\includegraphics[width=0.9\linewidth]{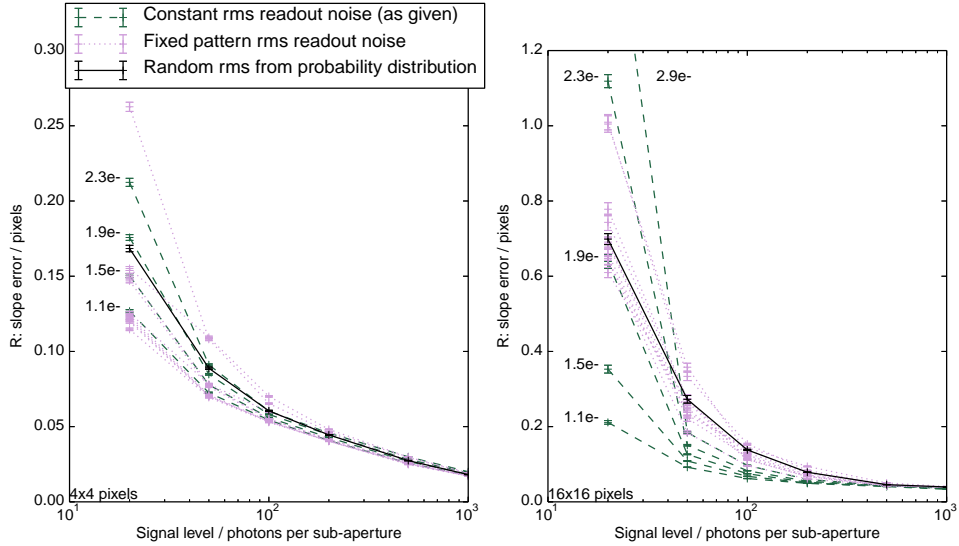}
\end{center}
\caption{A figure showing slope estimation error as a function of
  signal level for different sCMOS readout noise models, for a
  $4\times4$ (left) and $16\times16$ (right) pixel sub-aperture.  The
  dashed lines show slope error predicted using a constant rms readout
noise, the value of which is given adjacent to the lines.  The solid
line shows the average expected slope area from a sCMOS detector (the
``sCMOS Random'' model).  The dotted lines represent the ``sCMOS
Fixed'' model, and are shown to give an idea of spread in
performance.  }
\label{fig:scmosnoise}
\end{figure}

Fig.~\ref{fig:scmosnoise} compares slope estimation error ($R$) for
different readout noise models.  These include the ten ``\scmos
Fixed'' models identified earlier (e.g. Fig.~\ref{fig:noisepattern}),
the ``\scmos Random'' model, and also models with a range of constant
rms readout noise values.  By comparing the ``\scmos Random'' model
with the closest constant rms model for a given signal level, we can
get a feel for the {\em effective} readout noise of the detector for
that particular case.

\begin{figure}\begin{center}
\includegraphics[width=0.6\linewidth]{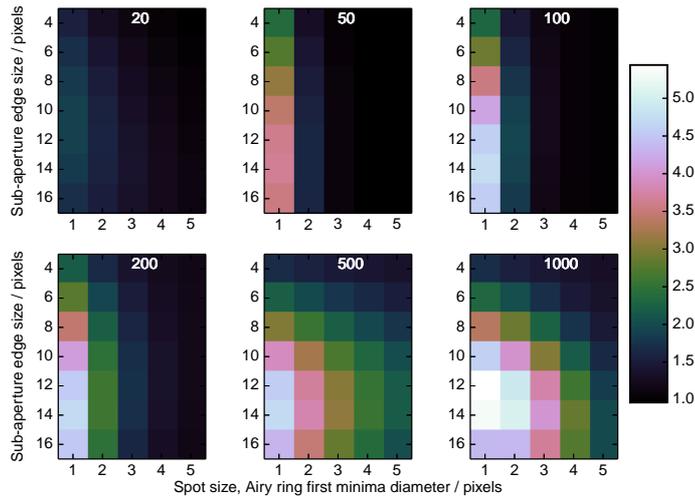}
\end{center}
\caption{A figure showing the effective Gaussian sCMOS rms readout noise that
  predicts the same slope estimation performance as that predicted by
  random sampling of the rms readout noise probability distribution.
  Values are given as a function of sub-aperture size and spot
  diameter (Airy ring first minimum diameter).  Six different signal
  levels are shown, with the number of photons per sub-aperture given
  inset in the figures.}
\label{fig:scmosnoise2d}
\end{figure}

To make sense of this information, and to provide a useful reference
for future instrument modelling, Fig.~\ref{fig:scmosnoise2d} shows the
single-value rms readout noise that will provide the same performance
as predicted by the ``\scmos Random'' model, for different
sub-aperture and spot sizes.  To use this figure when modelling a
specific \ao instrument, the known sub-aperture and spot size can be
used to read off an effective rms Gaussian readout noise for a \scmos
detector on the figure, i.e.\ a single readout value for the detector.
This effective rms Gaussian readout noise can then be used to predict \ao
system performance, giving a similar result as that expected if the full
randomly sampled rms readout noise probability distribution had been
used, but with reduced complexity.


It is important to note that different \scmos detector generations and
chip sizes will have a different rms readout noise probability
distribution.  We therefore recommend that an equivalent to
Fig.~\ref{fig:scmosnoise2d} should be generated for the specific
detector family under consideration in an instrument design.  Using
this information will then allow a more accurate prediction of
instrumental performance to be made.

\subsubsection{Elongated spots for laser guide stars}
\begin{figure}\begin{center}
\includegraphics[width=0.6\linewidth]{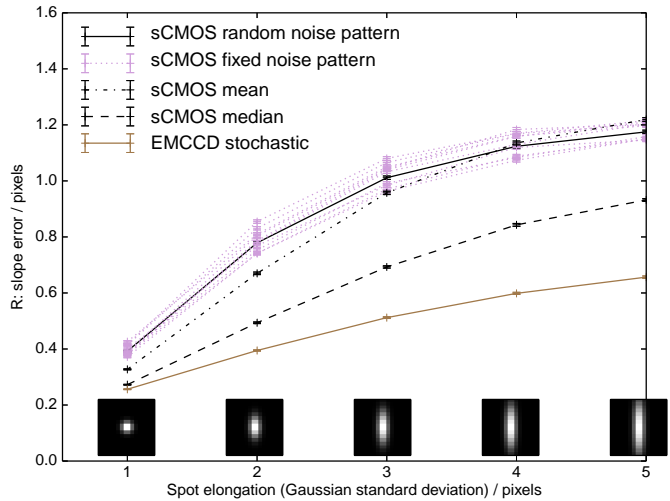}
\end{center}
\caption{A figure showing slope estimation error as a function of LGS
  elongation, with the relevant elongated spot PSFs shown inset
  ($16\times16$ pixel sub-apertures).  The different detector readout
  noise models are given in the legend.  Ten different fixed-pattern
  sCMOS rms readout noise patterns are shown undifferentiated, to give
  an idea of the spread in performance due to the random nature of the
  readout noise probability distribution.}
\label{fig:lgs}
\end{figure}

So far, we have only considered Shack-Hartmann \psfs with circular
symmetry.  However, it is also important to consider the case when
extended \lgs sources are used, producing elongated \psfs.
Fig.~\ref{fig:lgs} shows slope estimation error as a function of
elongation for a $16\times16$ sub-aperture.  A two-dimensional
Gaussian model has been used for the \lgs spot \psf, with the Gaussian
standard deviation in one dimension investigated.  We note that using
a standard deviation of unity gives a spot of size broadly equivalent
to an Airy disk with the diameter of the first minimum being six
pixels.  We apply the different readout noise models to these
elongated spots as described previously.

It can be seen that \emccd technology provides lowest error.  All
three \emccd readout noise models predict very similar performance,
and so only the ``\emccd Stochastic'' model is shown for clarity.  It
is interesting to note that as the spot becomes more elongated, \scmos
performance predicted using the ``\scmos Mean'' model becomes closer
to that predicted by the ``\scmos Random'' model, thus suggesting that
a simple model for \scmos readout noise is applicable for elongated
\lgs spots.  We note that in a real \ao system, the degree of \lgs
elongation will depend on sub-aperture position within the telescope
pupil, and some sub-apertures will remain almost un-elongated.  In
this case, the simple \scmos readout noise model is optimistic, and so
we recommend that a full \scmos readout noise model based on the rms
readout noise probability distribution should be used whenever readout
noise is a key instrument design consideration.

\subsubsection{Considerations of quantum efficiency}
\begin{figure}\begin{center}
\includegraphics[width=0.6\linewidth]{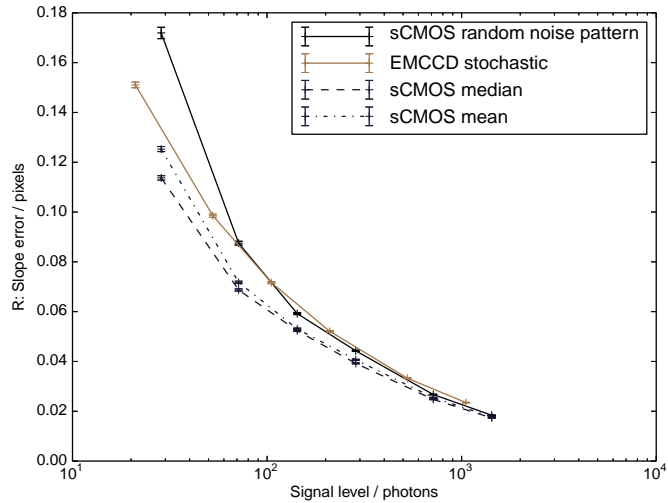}
\end{center}
\caption{A figure showing slope estimation error as a function of
  incident signal level for a $4\times4$ pixel sub-aperture.  A 70\%
  QE is assumed for sCMOS models, and 95\% QE for EMCCD models.  }
\label{fig:scaledForQe}
\end{figure}

We have so far ignored detector \qe, and assumed identical \qe for all
detector models (though we halve the effective \qe for the simple
\emccd model).  The \qe of \emccd devices can reach 95\% (e.g.\ the
Andor iXon3), while for \scmos detectors, it is closer to 70\%
(e.g.\ the Andor Zyla 4.2).  Fig.~\ref{fig:scaledForQe} shows slope
estimation error once \qe is taken into account, and can be compared
directly with Fig.~\ref{fig:scmosres4} (which assumes identical \qe).
It can be seen here that \emccd performance is now at least as good as
that predicted by the ``\scmos Random'' model, i.e.\ that in practice,
an \emccd detector is likely to perform as well as a \scmos detector
for $4\times4$ pixel sub-apertures (and, as we have seen previously,
better for larger sub-apertures).

\subsection{Astrometric accuracy}
\label{sect:astrometry}
We have investigated the effect of detector readout noise models on
image centroiding accuracy.  In addition to importance for \ao
systems, accurate position determination is critical for astrometric
techniques.  

We have shown that there are only small differences in estimation
accuracy between the commonly used simple \emccd model, and a full
stochastic gain mechanism model.  However, for some astrometric
observations, this difference may be critical, and therefore we
recommend that the full stochastic gain mechanism (or the \emccd
output probability distribution model, which is almost identical)
should be used, until it can be demonstrated that the simple model is
sufficient for each particular instrument study.  We note here that
the stochastic model is computationally more expensive than other models.

When using \scmos technology for astrometric applications, greater
care is required.  We have shown that a simple model of \scmos readout
noise based on a single rms value for all pixels (whether the median
or mean) is optimistic.  Therefore, a model for \scmos readout noise
that uses the per-pixel probability distribution for rms noise, is
essential.  Further model improvements can be made if the precise rms
readout noise pattern for a physical detector under consideration can
be used (i.e.\ once the detector has been acquired), though we do not
consider this further here.

\section{Conclusions}
We have investigated detector readout models for \scmos and \emccd
technologies, and the effect that these models have on slope
estimation accuracy for Shack-Hartmann wavefront sensors used in \ao
systems.  Our findings are also relevant to any problem involving
image centre of mass location, including astrometry.  We find that in
general \emccd technology offers better performance than \scmos
technology for Shack-Hartmann wavefront sensors and other applications
requiring centre of mass calculations.  

We find that the commonly used simple model for \emccd readout
(halving the effective \qe and assuming a sub-electron readout noise)
is sufficient for \ao applications with predicted slope estimation
accuracy differing only slightly from when using a full Monte-Carlo
stochastic gain mechanism model.  A model based on \emccd probability
output distribution also performs almost identically to the stochastic
gain model.

For \scmos technology, we find that the commonly used model that uses
a single rms readout value for all pixels (whether the median or mean)
produces optimistic results, which can predict better performance than
that obtained by \emccd detectors.  However, more reliable performance
estimates during instrument development and design studies can be made
by taking a typical \scmos rms readout noise probability distribution
into account, and we find that this model generally predicts worse
performance than that obtained by \emccd detectors.  Ideally, many
random samples of this distribution should be taken, so that an
average (and worst case) performance estimate for \scmos technologies
can be obtained.  A key finding is that using the median or mean
\scmos rms readout noise value is not sufficient to accurately predict
instrumental performance: the full probability distribution for \scmos
readout noise should be used.

\acknowledgments
This work is funded by the UK Science and Technology Facilities
Council, grant ST/I002871/1 and ST/L00075X/1.

%\bibliography{mybib}
\bibliographystyle{spiejour}

\vspace{2ex}\noindent{\bf Alastair Basden} has extensive expertise in low noise
detectors and adaptive optics, including real-time control and
simulation.  He is an eternal postdoc at Durham University.

%\listoffigures
%\listoftables

\end{document}